# Mitigation of Gilbert Damping in the CoFe/CuO$_x$ Orbital Torque System

*Shilei Ding\*, Hanchen Wang, William Legrand, Paul Noël, Pietro Gambardella\**

Department of Materials, ETH Zürich, 8093 Zürich, Switzerland

ABSTRACT: Charge-spin interconversion processes underpin the generation of spin-orbit torques in magnetic/nonmagnetic bilayers. However, efficient sources of spin currents such as 5$d$ metals are also efficient spin sinks, resulting in a large increase of magnetic damping. Here we show that a partially-oxidized 3$d$ metal can generate a strong orbital torque without a significant increase in damping. Measurements of the torque efficiency ξ and Gilbert damping α in CoFe/CuO$_x$ and CoFe/Pt indicate that ξ is comparable. The increase in damping relative to a single CoFe layer is $\Delta\alpha < 0.002$ in CoFe/CuOx and $\Delta\alpha \approx 0.005 - 0.02$ in CoFe/Pt, depending on CoFe thickness. We ascribe the nonreciprocal relationship between $\Delta\alpha$ and $\xi$ in CoFe/CuO$_x$ to the small orbital–to–spin current ratio generated by magnetic resonance in CoFe and the lack of an efficient spin sink in CuO$_x$. Our findings provide new perspectives on the efficient excitation of magnetization dynamics via the orbital torque.

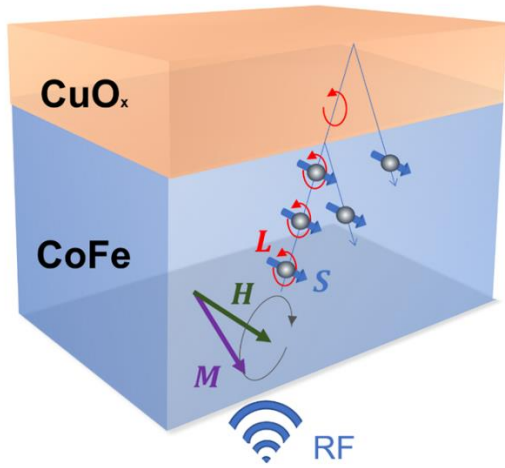 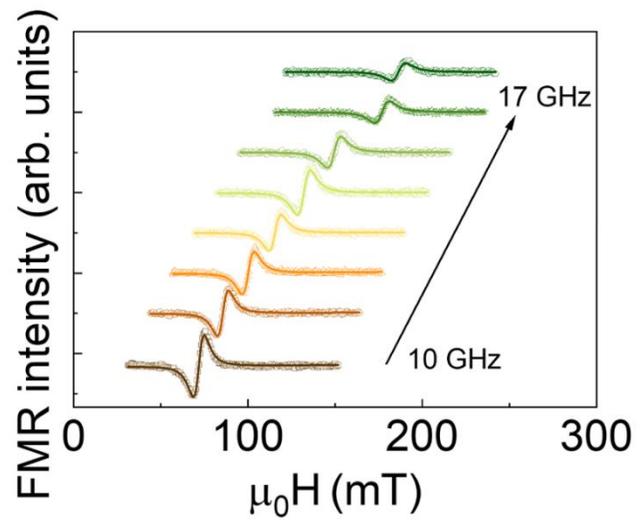

Magnetic damping determines the relaxation rate and energy dissipation associated with magnetization dynamics.[1-3] The phenomenological Gilbert damping parameter $\alpha$ can be separated in two parts, an intrinsic damping arising mainly from the spin-orbit interaction in a ferromagnet[4,5] and an extrinsic damping arising mainly from two-magnon scattering,[6] spin pumping,[7,8] and spin memory loss at interfaces.[9-11] These additional processes are prominent in metallic thin films and multilayers, where they contribute to the relaxation rates and enhance the damping of the magnetic materials. Reducing the extrinsic damping in such systems is of primary importance for spintronic applications that rely on energy-efficient current-induced magnetic oscillations,[12] switching,[13-15] and long-distance magnon propagation.[16]

Spintronic devices consisting of a ferromagnetic layer (FM) adjacent to a nonmagnetic layer (NM) offer multiple ways to manipulate and detect magnetization via charge-spin interconversion processes.[14,17,18] Owing to spin-orbit coupling (SOC), the injection of an electric current in an FM/NM bilayer results in the generation of a spin current transverse to the charge flow, which is eventually absorbed by the magnetization of the FM, generating a spin-orbit torque (SOT).[14,19] The absorption of a spin current with a polarization that is aligned with the magnetization offsets the intrinsic dissipation of angular momentum, resulting in an increase or decrease of the damping depending on the sign of the spin current.[20,21] The reciprocal of this process is spin pumping, which consists of exciting the magnetization dynamics by a time-dependent magnetic field and the consequent dissipation of angular momentum in the form of a spin current flowing from the FM into the NM. This spin current can ultimately be converted into a charge current by SOC.[22,23]

The NM thus acts either as a spin source or spin sink owing to the charge-spin interconversion processes promoted by SOC, as exemplified by the direct and inverse spin Hall effect.[24] Heavy metals such as Pt and W are among the most efficient sources but also sinks of spin currents due to the large intrinsic SOC of the 5*d* bands crossing the Fermi energy. Applications that rely on the excitation of magnetization dynamics by SOT would therefore greatly benefit from the development of alternative SOT sources that do not involve a concomitant increase of damping through extrinsic mechanisms.

Recently, the study of SOT has been extended to include light metals and their oxides as sources of orbital angular momentum.[25-40] In such systems, the orbital Hall or the orbital Rashba Edelstein effect promotes the accumulation of nonequilibrium orbital momenta at the NM interface, which then propagate into the FM as a diffusive orbital.[41-47] Although the generation of an orbital current does not require SOC in the NM, the conversion to an orbital torque acting on the magnetization relies on SOC in the FM to efficiently couple the orbital current to the local spins of the FM.[26-28] The inverse process of orbital pumping has also been investigated, whereby the excitation of resonant magnetization dynamics in a ferromagnet leads to the dissipation of an orbital current and conversion into a charge current in a nearby nonmagnetic layer.[48-56] Light transition metals and their oxides have been predicted to display orbital Hall conductivities similar to or even larger compared to the spin Hall conductivity of heavy metals,[45,46,56,57] which significantly expands the number of materials available for efficient charge-angular momentum interconversion processes. Moreover, the weak SOC of the light metals makes them poor spin sinks, leading to spin backflow in the FM. The question therefore arises on whether FM/NM bilayers including an NM light-metal exhibit weaker damping compared to heavy metals while still providing large SOT efficiencies.

In this work, we report on the orbital torque efficiency and damping of 3*d* FM/NM bilayers and compare them with that of FM/Pt and single FM layers, as exemplified in Figure 1(a). We take a partially-oxidized Cu layer, hereafter indicated as $CuO_x$, as an NM light-metal system in which the orbital Rashba-Edelstein effect induces orbital accumulation at the interface of an FM layer, leading to strong orbital torques comparable to heavy metals.[30-35,58,59] As FM, we take $Co_{25}Fe_{75}$, henceforth indicated as CoFe, owing to its reduced damping compared to other metallic FM.[60] We performed harmonic Hall voltage measurements of the SOTs and ferromagnetic resonance (FMR) measurements to obtain the Gilbert damping of CoFe/CuO$_x$ and CoFe/Pt samples, comparing $\xi_{DL}^E$ and $\alpha$ as a function of the thickness of the CoFe layer. Our results indicate that light metal systems can provide large orbital torques together with a lower Gilbert damping compared to heavy metals.

The samples were grown on Si/SiO$_2$ substrates by dc magnetron sputtering. We deposited first a buffer SiN(6) layer (thickness in nanometers) to prevent interfacial oxidation followed by CoFe($t_{CoFe}$)/X bilayer where X= CuO$_x$(5), Pt(5), Cu(5)/SiN(6), and SiN(6). Here CuO$_x$(5) stands for a 5-nm-thick Cu layer naturally oxidized in air for 2 days before the measurements. Consistently with previous work, CuO$_x$(5) is only partially oxidized from the top, enabling a substantial flow of current in a region with a decreasing oxygen concentration towards the FM layer, as required for the generation of an orbital torque.[30-35] The magnetization of the different layers was measured by a superconducting quantum interference device. Atomic force microscopy evidenced a root mean square roughness of the samples of less than 0.2 nm, indicating excellent film homogeneity (See Supporting Information Note S1). The electrical transport and SOT measurements were carried out on Hall bar devices with a width of 10 μm patterned by photolithography and lift-off, using an alternate (ac) excitation current with a frequency of 10 Hz, see Figure 1(b). The first and second harmonic longitudinal and transverse resistances were measured simultaneously to extract the SOT.[61,62] Details of the torque measurements are given in Supporting Information Note S2-S5. The FMR spectra were recorded using a broadband field-modulated setup, with measurements performed directly on the full films flipped onto a coplanar waveguide. The Gilbert damping parameter was extracted from the standard linewidth analysis of the resonance peaks against the field at different frequencies $\omega/2\pi = 10$ to $17$ GHz.[60,63] All measurements were performed at room temperature.

Figure 1(c) shows the effective magnetic field $\mu_0 H_{DL}$ corresponding to the damping-like SOT as a function of electric field $E$ applied to the Hall bars of CoFe/Pt, CoFe/CuO$_x$, and CoFe/Cu. As expected, $\mu_0 H_{DL}$ is very small in CoFe/Cu due to the absence of significant charge-spin and charge-orbital conversion in Cu, whereas it increases linearly with $E = IR/L$ in CoFe/Pt and CoFe/CuO$_x$, where $I$ is the electric current, $R$ the longitudinal resistance and $L$ the length of the Hall bar. In agreement with previous works, we assign the damping-like SOT of CoFe/Pt to the dominant spin Hall effect of Pt and that of CoFe/CuO$_x$ to the orbital Rashba-Edelstein effect of the Cu/CuO$_x$ interface in the partially-oxidized CuO$_x$ layer.[30-35,58,59] By fitting the data with a linear function we obtain the damping-like torque efficiency per unit applied electric field

$$\xi_{DL}^{E} = \frac{\frac{2e}{\hbar} M_s t_{CoFe} \mu_0 H_{DL}}{E}, \qquad (1)$$

where $e$ is the electric charge, $\hbar$ the reduced Planck constant and $M_s$ and $t_{CoFe}$ are the saturation magnetization and thickness of the CoFe layer, respectively. Note that $M_s t_{CoFe}$ is the areal saturation magnetization, which is not affected by a possible magnetic dead layer in CoFe (See Supporting Information Note S3). The fits give $\xi_{DL}^{E} = 1.03 \pm 0.05 \times 10^5\ \Omega^{-1}\text{m}^{-1}$ for CoFe(3)/CuO$_x$(5) and $\xi_{DL}^{E} = 2.6 \pm 0.1 \times 10^5\ \Omega^{-1}\text{m}^{-1}$ for CoFe(3)/Pt(5), in agreement with previous works.[33]

Figure 1(d) shows $\mu_0 H_{DL}$ as a function of $t_{CoFe}$ for different NM layers. In a typical spin torque system such as CoFe($t_{CoFe}$)/Pt(5), $\mu_0 H_{DL}$ decreases as $t_{CoFe}^{-1}$, as expected for a spin current absorbed within a spin-dephasing length ($\lambda_D$) of the interface and the effective field is diluted in the thick FM layers. In transition-metal FM, $\lambda_D \approx 1$ nm and the spin torque is normally regarded as an interfacial effect. In contrast, $\mu_0 H_{DL}$ in CoFe($t_{CoFe}$)/CuO$_x$(5) increases up to 4 nm before decreasing as $\sim t_{CoFe}^{-1}$. This behavior is in line with that expected of an orbital torque, as the nonequilibrium orbital momenta injected from CuO$_x$ into CoFe do not couple directly to the magnetization and have to be converted to a spin moment first in order to exert a torque.[44] The orbital-to-spin conversion process typically saturates over a thickness of a few nanometers in an FM layer.[27,28,36]

We now discuss the Gilbert damping in these bilayers. Figures 2(a,b) show the field derivatives of the real and imaginary microwave transmission coefficient $S_{21}$ of CoFe(3)/CuO$_x$(5) measured in the coplanar waveguide as a function of the in-plane magnetic field at various excitation frequencies. The FMR curves show a non-monotonic change of amplitude with frequency, caused by an imperfect impedance matching at the sample location. This variation, however, does not impact the subsequent determination of line widths and damping parameters. The FMR field $\mu_0 H_{res}$ increases as a function of $\omega$ as shown in Figure 2(c). By fitting the data to the Kittel equation $\omega^2 = \gamma^2(\mu_0 H_{res} - \mu_0 H_k)(\mu_0 H_{res} - \mu_0 H_k + \mu_0 M_{eff})$, where $\gamma = 2\pi \left(\frac{g_{eff}}{2}\right) \times$ 28 GHz/T is the gyromagnetic ratio, $\mu_0 H_k$ is the anisotropic field, and $\mu_0 M_{eff}$ represents the uniaxial anisotropy perpendicular to the film. We obtain the effective $g$-factor $g_{eff} = 2.0 \pm 0.2$, effective magnetization $\mu_0 M_{eff} = 1.50 \pm 0.01$ T ($M_{eff} = (1.20 \pm 0.01) \times 10^3$ kA/m) and anisotropy field $\mu_0 H_k = 2 \pm 3$ mT. An analogous set of measurements for CoFe(3)/Cu(5)/SiN(6) and CoFe(3)/Pt(5) yields $g_{eff} = 2.0 \pm 0.3$ for both samples. A Lorentzian fit of the FMR spectra was used to extract the full-width at half-maximum of the FMR peak, $\mu_0 \Delta H$. Figure 2(d) shows the frequency dependence of $\mu_0 \Delta H$ of CoFe with different NM layers, which follows the linear relationship

$$\mu_0 \Delta H = \mu_0 \Delta H_0 + 2\alpha\omega/\gamma, \qquad (2)$$

where $\mu_0 \Delta H_0$ is the inhomogeneous linewidth broadening.[63] Linear fits to the data as a function of $\omega$ yield $\alpha$.

Figure 3(a) reports the values of $\alpha$ obtained in different sample series as a function of $t_{CoFe}$. Our data show that the CoFe/Pt bilayers present the largest damping of all samples over the entire range of thickness of CoFe. We measure $\alpha = 0.034 \pm 0.005$ in CoFe(3)/Pt(5), whereas $\alpha = 0.016 \pm 0.002$ in CoFe(3)/Cu(5)/SiN(6) and $\alpha = 0.013 \pm 0.001$ in CoFe(3)/CuO$_x$(5). This difference originates from the contribution of spin pumping to the damping in CoFe(3)/Pt(5), in which Pt is a very efficient spin sink, and, to a much smaller extent, to the damping of the proximity-induced magnetization in Pt. The latter point is validated by measuring a CoFe(3)/Cu(1)/Pt(5) layer in which the Cu spacer minimizes the proximity effects between CoFe and Pt, for which $\alpha = 0.028 \pm 0.003$ (not shown). The comparison between the damping of CoFe(3)/CuO$_x$(5) and CoFe(3)/Pt(5) shows that an orbital torque system leads to a much smaller enhancement of the damping $\alpha$ relative to a spin torque system in Pt, comparable to that of CoFe(3)/Cu(5)/SiN(6), i.e., a magnetic layer that is not coupled to an efficient source of angular momentum.

Further insights can be obtained by analyzing the dependence of $\alpha$ on $(M_s t_{CoFe})^{-1}$, as shown in Figure 3(b). In the spin Hall scenario, the total damping can be written as $\alpha = \alpha_0 + \Delta\alpha$,[8,64,65] where $\alpha_0$ is the damping without the spin sink. The damping enhancement $\Delta\alpha$ is associated to the spin-pumping contribution following the relation

$$\Delta\alpha = \frac{\gamma\hbar}{4\pi M_s t_{CoFe}} g^{\uparrow\downarrow}, \qquad (3)$$

where $g^{\uparrow\downarrow}$ is the spin-mixing conductance. In this expression, within the spin Hall scenario, we assumed that the thickness of the nonmagnetic layer $t_N$ is larger than the spin diffusion length $\lambda_N$, thus neglecting spin backflow, and considering only the real part of $g^{\uparrow\downarrow}$. Generally, other contributions to the damping can appear due, e.g., to scattering at interfacial defects and two-magnon scattering, which is expected to vary as $t_{CoFe}^{-2}$.[65] In the CoFe/Pt series, the damping decreases as $t_{CoFe}^{-1}$, consistently with the spin pumping contribution given by Eq. (3). A fit of $\alpha$ as a function of $(M_s t_{CoFe})^{-1}$ gives $g^{\uparrow\downarrow} = 32 \pm 4$ nm$^{-2}$, comparable to values reported for Pt interfaces,[64] and $\alpha_0 = 0.024 \pm 0.001$, much larger than the values of $\alpha$ observed for the samples that do not contain Pt. Most importantly, the behavior of $\alpha$ in the CoFe/CuO$_x$ series is remarkably different from that of CoFe/Pt and typical spin pumping systems and similar to that of CoFe/Cu, which has negligible torques. Within the experimental scattering of the data, $\alpha$ of CoFe/CuO$_x$ keeps a relatively constant value over the entire range of $t_{CoFe}$. Moreover, $\alpha_0$ is smaller than in CoFe/Pt, which we attribute to the smaller interfacial losses of CoFe/CuO$_x$ relative to CoFe/Pt.[66] The single-layer CoFe series capped by SiN has the lowest damping, likely due to the weak spin memory loss at the CoFe/SiN interface and the absence of other conducting layers. In this series, the lower damping makes it possible to appreciate an increase of $\alpha$ at low thickness compatible with two-magnon scattering.[65]

Figure 4 compares the two quantities associated with the injection (dissipation) of angular momentum in CoFe from (to) either Pt or CuO$_x$, namely the SOT efficiency $\xi_{DL}^E$ and the increase of damping $\Delta\alpha$. Figure 4(a) shows that $\xi_{DL}^E$ is already close to saturation for $t_{CoFe} = 1$ nm in the CoFe/Pt series, whereas $\xi_{DL}^E$ saturates at $t_{CoFe} = 4$ nm in the CoFe/CuO$_x$ series. At this thickness, the orbital torque efficiency reaches $\xi_{DL}^E = 1.7 \times 10^5$ $\Omega^{-1}$m$^{-1}$, which is 65 % of $\xi_{DL}^E$ in CoFe/Pt. Figure 4(b) shows the increase of damping *vs.* $t_{CoFe}$ calculated as $\Delta\alpha(t_{CoFe}) = \alpha(t_{CoFe}) - \alpha_0$. Here the difference between the two samples series is striking, as $\Delta\alpha$ ranges from 0.02 to 0.005 in CoFe/Pt and is $\approx 0$ in CoFe/CuO$_x$. Therefore, we conclude that the increase of damping $\Delta\alpha$ due to angular momentum pumping is much smaller in an orbital torque system like CoFe/CuO$_x$ compared to a spin torque system like CoFe/Pt. This result can be rationalized by considering that the precession of the magnetization in the ferromagnetic layer excites both a spin current and an orbital current.[49-51] The partially oxidized Cu layer does not provide an efficient sink of the spin current. On the contrary, a significant spin backflow is present, which minimizes the increase of damping, as sketched in Figure 1(a). The orbital current generated by CoFe, on the other hand, is at least one order of magnitude smaller than the spin current, consistently with the ratio of orbital and spin magnetization in a transition-metal FM.[67] As a consequence, orbital pumping can be expected to provide a much smaller source of damping compared to spin pumping. Assuming that the dissipated angular momentum by orbital pumping in CoFe/CuO$_x$ is less than 10% of the spin pumping in CoFe/Pt,[49] we expect $\Delta\alpha \approx 0.001$ in CoFe/CuO$_x$, which is in line with our results and smaller than the measurement error in our experiments.

In pure spin current systems, the relationship between SOT and the damping contribution due to spin pumping has been investigated both theoretically [68-70] and experimentally.[10,71,72] In the standard drift-diffusion approach[68,73] and neglecting spin backflow, the damping-like SOT efficiency per unit of electric field is given approximately by $\xi_{DL}^{E} = \frac{2e^2}{h} g^{\uparrow\downarrow} \lambda_N \rho_N \sigma_S$,[74] where $\lambda_N$, $\rho_N$, and $\sigma_S$ are the spin diffusion length, resistivity, and spin Hall conductivity of the NM layer, respectively. By further considering Eq. (3), the ratio $\xi_{DL}^{E}/\Delta\alpha$ turns out to be proportional to $\lambda_N \rho_N \sigma_S M_s t_{CoFe}$ (See Supporting Information Note S6). For a fixed thickness of the NM layer, we expect $\xi_{DL}^{E}/\Delta\alpha \sim M_s t_{CoFe}$, which is indeed verified for CoFe/Pt (See Supporting Information Note S6). Whether such a relationship has an equivalent orbital counterpart remains to be ascertained, together with the role played by the orbital mixing conductivity. Our data show that $\Delta\alpha \approx 0$ independently of thickness and for relatively high values of $\xi_{DL}^{E}$. For this reason, determining the ratio $\xi_{DL}^{E}/\Delta\alpha$ in CoFe/CuO$_x$ is challenging. As $\Delta\alpha$ is close to 0, the ratio $\xi_{DL}^{E}/\Delta\alpha$ is not well defined. An alternative method is to compute $\delta\alpha = \alpha(\text{CoFe/X}) - \alpha(\text{CoFe/SiN})$ and compare the ratio $\xi_{DL}^{E}/\delta\alpha$ for X = Pt and X = CuO$_x$. As we show in Figure 4(c) and (d), $\xi_{DL}^{E}/\delta\alpha$ is indeed larger in the CoFe/CuO$_x$ series compared to CoFe/Pt. Therefore, we conclude that it is possible to induce strong torques without a concomitant increase in damping.

Before concluding, we note that further insights into charge-orbital interconversion processes are desirable. Open questions concern the transmission of orbital currents at interfaces, as exemplified by an orbital mixing conductance parameter analogous to $g^{\uparrow\downarrow}$, the interfacial loss of orbital angular momentum, and the local vs global Onsager reciprocity of the charge-orbital interconversion.[75,76]

In summary, we reported on the combination of efficient orbital torque and low increase in the Gilbert damping in CoFe/CuO$_x$ bilayers. While the orbital Rashba-Edelstein effect in CuO$_x$ provides an efficient source of orbital angular momentum that is converted into torque in CoFe, the reciprocal process of damping of the FMR-induced orbital and spin currents flowing from CoFe into CuO$_x$ appears to be less efficient. We attribute the results to the small orbital-to-spin magnetization ratio of the transition-metal FM, which results in the generation of proportionally smaller orbital currents flowing out of CoFe compared to spin currents. The analysis of the angular momentum pumping contribution shows that $\Delta\alpha \approx 0$ for CoFe/Cu and CoFe/CuO$_x$. The absolute increase of damping relative to a single CoFe layer is also significantly smaller in CoFe/CuO$_x$ ($\delta\alpha = 0.003$ to $0.008$) than in CoFe/Pt ($\delta\alpha = 0.028$ to $0.017$), particularly for $t_{\text{CoFe}} < 5$ nm. The ratio $\xi_{DL}^{E}/\delta\alpha$ is thus higher in CoFe/CuO$_x$ relative to CoFe/Pt, in which Pt is both an efficient source and sink of spin currents. Our results provide insight into the relationship between orbital currents and damping in FM/NM heterostructures and are promising for the implementation of orbital torque oscillators with reduced damping compared to spin torque oscillators.

FIGURES

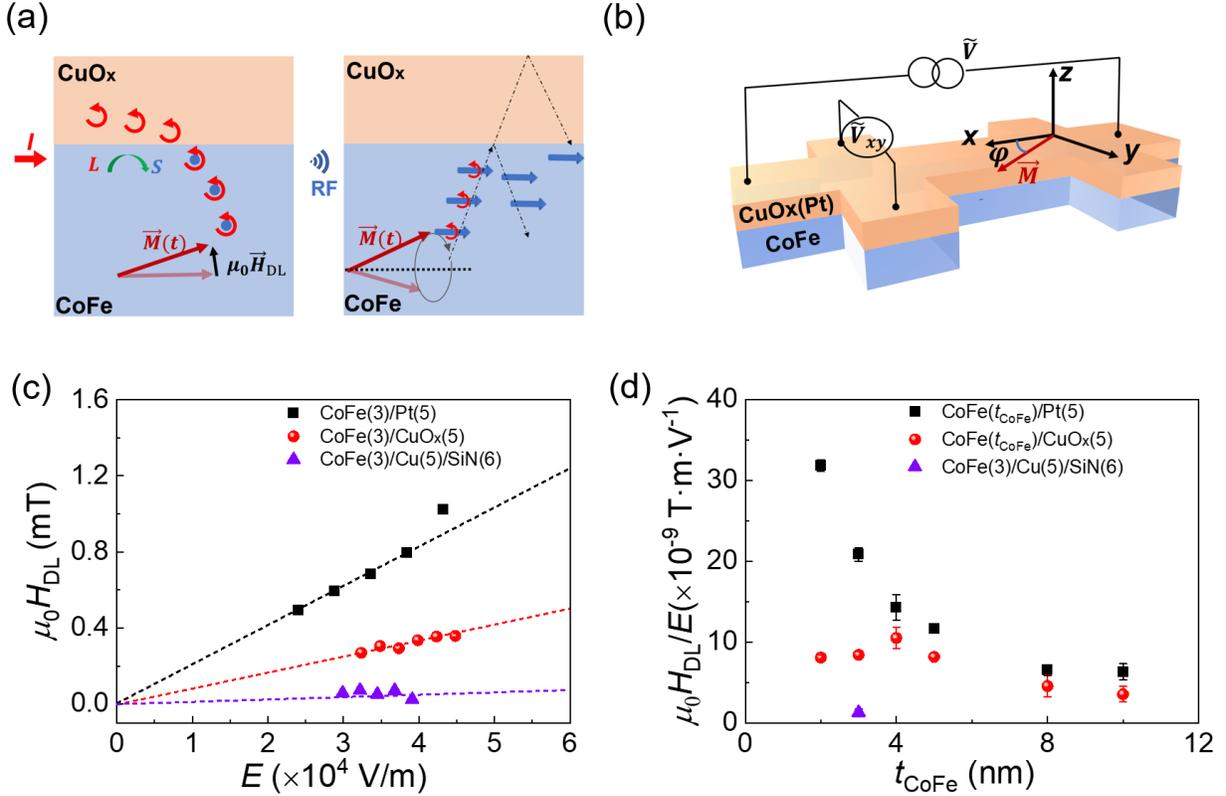

**Figure 1** (a) Schematics of current-induced orbital torques and orbital pumping. Left: The nonequilibrium orbital angular moment generated by an electric current in $CuO_x$ due to the orbital Rashba-Edelstein effect (*L*, circular arrows) diffuses into the CoFe layer, where it is converted into a spin moment (*S*, dot in the circular arrows) due to the spin-orbit coupling of CoFe. Right: The precession of the magnetization in CoFe excites a spin and an orbital current that diffuse into $CuO_x$. The spin current is not absorbed by $CuO_x$, leading to spin backflow. (b) Schematic of the Hall bar and measurement configuration. (c) Effective damping-like spin-orbit field $\mu_0 H_{DL}$ in CoFe/Pt, CoFe/$CuO_x$, and CoFe/Cu/SiN as a function of applied electric field and (d) damping-like spin-orbit field $\mu_0 H_{DL}$ per electric field as a function of the thickness of CoFe.

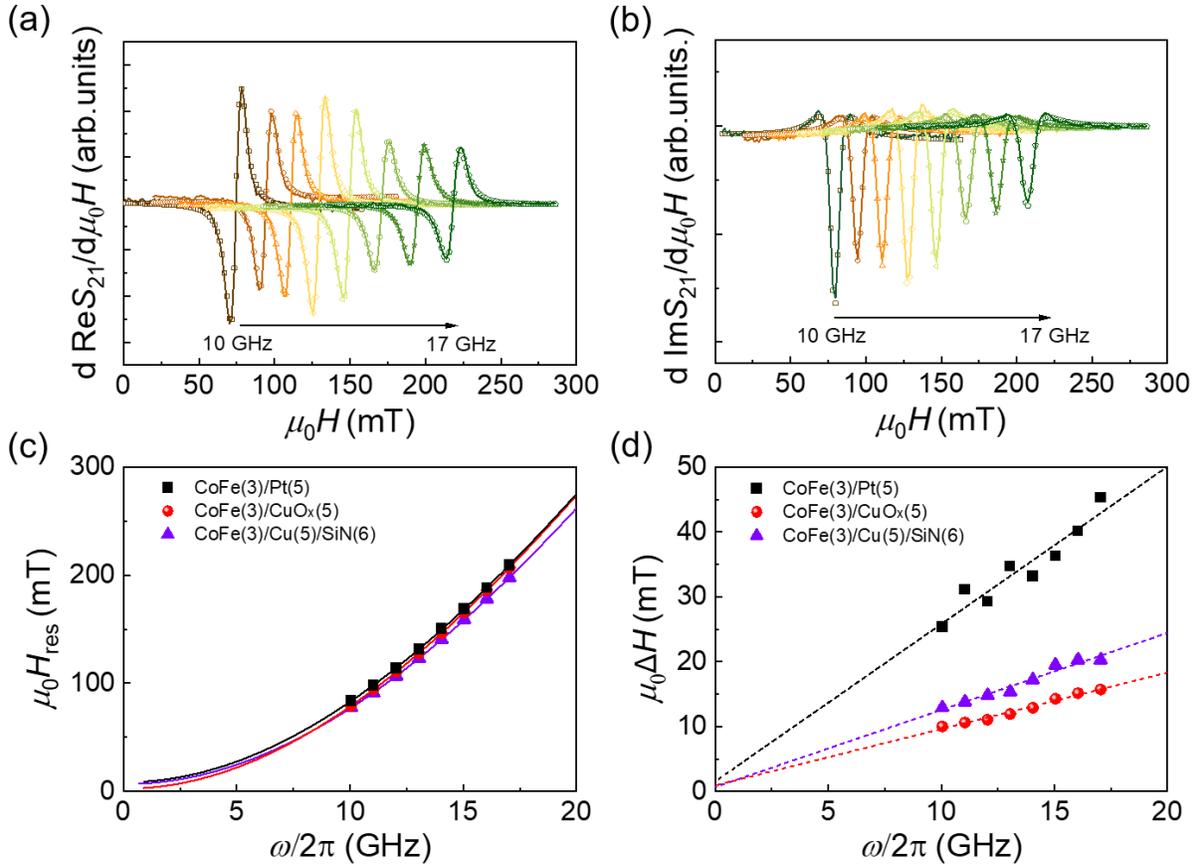

**Figure 2** (a) $d\,\text{Re}\,S_{21}/d\mu_0 H$ and (b) $d\,\text{Im}\,S_{21}/d\mu_0 H$, corresponding to the real and imaginary parts of the microwave transmission coefficient $S_{21}$ of the waveguide, collected at various excitation frequencies for CoFe(3)/CuO$_x$(5). (c) Frequency dependence of the resonance field $\mu_0 H_{\text{res}}$. The solid lines are the fits according to the Kittel equation. (d) Frequency dependence of the FMR linewidth $\mu_0 \Delta H$. The slope of the linear fits gives the Gilbert damping $\alpha$.

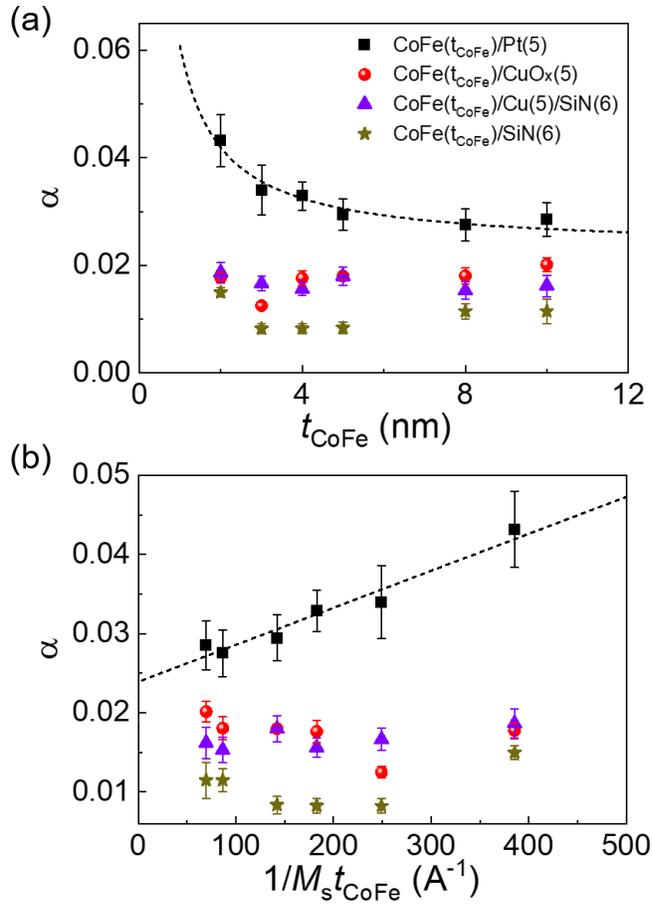

**Figure 3** (a) Thickness dependence of the Gilbert damping of CoFe with different NM layers. The dashed line is a fit by Eq.(3). (b) Same data as a function of $(M_s t_{\mathrm{CoFe}})^{-1}$. The dashed line is a linear fit.

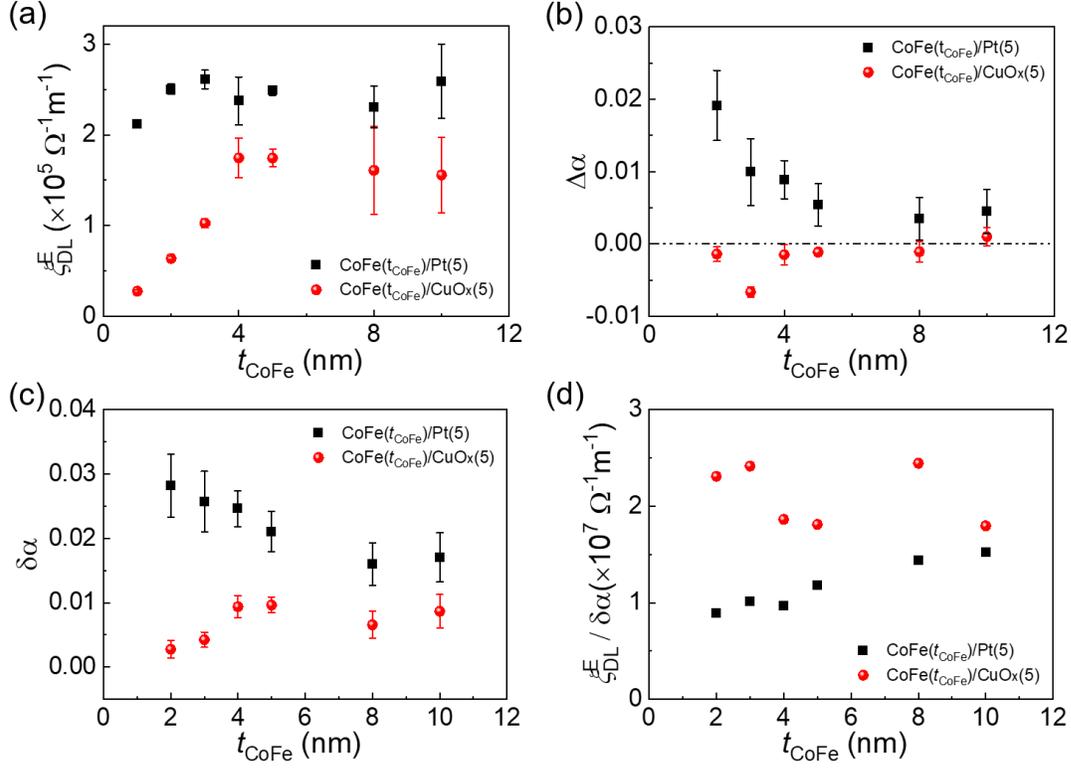

**Figure 4** (a) Damping-like spin-orbit efficiency $\xi_{DL}^E$ and (b) increase of damping $\Delta\alpha$ in CoFe/Pt and CoFe/CuO$_x$ as a function of the thickness of CoFe. The increase of damping $\Delta\alpha$ is defined as $\alpha - \alpha_0$, where $\alpha_0$ is the damping without the spin sink. $\alpha_0$ was determined from the linear fit of $\alpha$ vs $(M_s t_{CoFe})^{-1}$. (c) $\delta\alpha = \alpha(\text{CoFe/Pt}) - \alpha(\text{CoFe/SiN})$ and $\delta\alpha = \alpha(\text{CoFe/CuO}_x) - \alpha(\text{CoFe/SiN})$ in CoFe($t_{CoFe}$)/Pt(5) and CoFe($t_{CoFe}$)/CuO$_x$(5) as a function of CoFe thickness. (d) $\xi_{DL}^E/\delta\alpha$ as a function of CoFe thickness in CoFe($t_{CoFe}$)/Pt(5) and CoFe($t_{CoFe}$)/CuO$_x$(5).

ASSOCIATED CONTENT

**Supporting Information.** Surface roughness of the samples, anomalous Hall measurement of CoFe(3)/CuOx(5), planar Hall measurement of CoFe(3)/CuOx(5), the saturation magnetization of the samples, the resistance of CoFe($t_{CoFe}$)/Pt(5) and CoFe($t_{CoFe}$)/CuOx(5), measurement of the torque efficiency by harmonic voltage method. CoFe thickness dependence of $\xi_{DL}^{E}/\Delta\alpha$. CoFe thickness dependence of the field-like torque.

**Data Availability.** The data underlying this study are openly available in ETH Research Collection at 10.3929/ethz-b-000686806.

AUTHOR INFORMATION

**Corresponding Author**

Shilei Ding - Department of Materials, ETH Zürich, 8093 Zürich, Switzerland.
Email:shilei.ding@mat.ethz.ch

Pietro Gambardella - Department of Materials, ETH Zürich, 8093 Zürich, Switzerland.
Email:pietro.gambardella@mat.ethz.ch

**Authors**

Hanchen Wang - Department of Materials, ETH Zürich, 8093 Zürich, Switzerland.

William Legrand - Department of Materials, ETH Zürich, 8093 Zürich, Switzerland.

Paul Noël - Department of Materials, ETH Zürich, 8093 Zürich, Switzerland.

**Author Contributions**

S. D. and P. G. conceived and designed the experiments. S. D. fabricated the samples. S. D. and P. N. performed the torque efficiency measurement. H.C.W and W. L. performed the damping measurement. S. D. and P. G. wrote the manuscript. All authors discussed the results and provided comments on the manuscript.

**Notes**

The authors declare no competing financial interest.

ACKNOWLEDGMENT

This work was partially funded by the Swiss National Science Foundation (Grant No. 200020_200465). H.W. acknowledges the support of the China Scholarship Council (CSC) under Grant No. 202206020091. W. L. and P. N. acknowledge the support of the ETH Zurich Postdoctoral Fellowship Program (21-1 FEL-48 and 19-2 FEL-61).

x...